\documentstyle[11pt,aaspp4]{article}
\begin{document}
%
%
\title{\huge Stochastic Biasing and Galaxy-Mass Density Relation in 
the Weakly Non-linear Regime}
%
%
%
%
\author{Atsushi Taruya and Jiro Soda} 
\affil{
Department of Fundamental Sciences, FIHS, Kyoto University, 
Kyoto, 606-8501, Japan \\
}
\authoraddr{}
%
%
%
%
%
%
%
%
%
\def\k{\mbox{\boldmath$k$}}
\def\x{\mbox{\boldmath$x$}}
\def\v{\mbox{\boldmath$v$}}
\def\cmg{\la\dg\ra|_{\dm}}
\def\cmm{\la\dm\ra|_{\dg}}
\def\varm{\la\dm^2\ra}
\def\varg{\la\dg^2\ra}
\def\dm{\delta_m}
\def\dg{\delta_g}
\def\Dm{\Delta_m}
\def\Dg{\Delta_g}
\def\la{\langle}
\def\ra{\rangle}
\def\sigm{\sigma_m}
\def\eps{\epsilon}
\def\lm{\lambda}
%
%
%
%
\begin{abstract}
It is believed that the biasing of the galaxies plays an 
important role for understanding the large-scale structure of the universe. 
In general, the biasing of galaxy formation could be stochastic. 
Furthermore, the future galaxy survey might allow us to explore 
the time evolution of the galaxy distribution. 
In this paper, the analytic study of the galaxy-mass 
density relation and its time evolution is presented within 
the framework of the stochastic biasing. In the weakly non-linear regime, 
we derive a general formula for the galaxy-mass density relation 
as a conditional mean using the Edgeworth expansion. 
The resulting expression contains the joint moments of the total mass and 
galaxy distributions. Using the perturbation theory, 
we investigate the time evolution of the joint moments and 
examine the influence of the initial stochasticity on the 
galaxy-mass density relation. The analysis shows that the galaxy-mass 
density relation could be well-approximated by the linear relation. 
Compared with the skewness of the galaxy 
distribution, we find that the estimation of the higher order moments using 
the conditional mean could be affected by the stochasticity. 
Therefore, the galaxy-mass density relation 
as a conditional mean should be used with a caution as a tool for 
estimating the skewness and the kurtosis. 
\end{abstract}
%
%
%
\section{Introduction}
\label{sec: intro}
%
%
%
%
The enormous data of the 
large scale galaxy distribution will be obtained in the future.  
It is expected that the physics in the early universe can be explored 
by using these data. 
According to the standard picture based on the 
gravitational instability, the primordial fluctuations 
produced during the inflationary stage 
have evolved into the large scale structure observed in the 
galaxy sky survey. 
To compare the theoretical prediction for the density fluctuation with 
the observation, we need the relation between the total mass and the 
galaxies. Due to the lack of our knowledge of the galaxy formation 
process, however, the statistical uncertainty 
between the galaxies and the total mass density 
arises. This uncertainty affects the determination of the cosmological 
parameters from the observation (\cite{hamilton}). 

Denoting the fluctuation of total mass as $\dm$ and that of galaxy 
distribution as $\dg$, the linear biasing relation 
\begin{equation}
  \label{linear-bias}
\dg=b\dm, 
\end{equation}
is frequently used in the literature. Here, 
$b$ is the biasing parameter. 
A naive extension of (\ref{linear-bias}) to 
the quasi non-linear regime is obtained by taking the expansion 
in powers of $\dm$: 
\begin{equation}
\dg=f(\dm)=\sum_{n=1}\frac{b_n}{n!}(\dm)^n.
  \label{nl-bias}
\end{equation}
The non-linear biasing parameters $b_n$ is observationally determined  
by several authors (\cite{FG93}, \cite{F94}, \cite{GF94}). 
However, we must note that there is an assumption behind 
equations (\ref{linear-bias}) and (\ref{nl-bias}) that 
the statistics of the galaxy distribution is completely characterized by 
that of the total mass $\dm$. In reality, the  galaxy formation 
process could be stochastic (\cite{cenI}). Stochastic biasing  
is a notion to treat this situation. Note that the deterministic 
biasing corresponds to a special case of the stochastic biasing. 
Stochastic property of the galaxy biasing is recently studied by many 
authors. Pen (1997) has discussed an observational method to determine 
the stochasticity. Scherrer \& Weinberg (1998) has studied the influences of 
the non-linearity and the stochasticity on the statistics for the two point 
correlation function. 

Due to the gravitational dynamics and galaxy formation process, the relation 
of the galaxies and the total mass is generally non-linear and time 
dependent. The recent observation at high redshift shows that 
the galaxies at $z\simeq 3$ are strongly biased. The evidence of 
the strong biasing at high redshift suggests that the time 
evolution of the biasing should be considered (\cite{peacock}). 
Furthermore, Magliocchetti {\it et al}. (1998) showed that the measurement 
of the angular correlation function of the radio galaxies obtained 
from the Faint Images of the Radio Sky at Twenty centimeters is 
consistently explained by taking into account the evolution of the biasing. 
The time evolution of the galaxy biasing 
has been firstly studied by Fry (1996). Tegmark \& Peebles (1998) extended it 
to the stochastic biasing and examined the linear evolution. 
In the previous paper (\cite{taruya99}), we have investigated 
the quasi non-linear evolution of the stochastic biasing. 

A general formalism for the stochastic biasing 
is recently proposed by Dekel \& Lahav (1998). The non-linearity and 
the stochasticity of the galaxy biasing are measured by the 
conditional mean and its scatter. 
  They performed the numerical 
  simulation and evaluated the conditional mean of $\dg$ at a given 
  total mass fluctuation $\dm$. Blanton {\it et al}. (1998) studied the 
  galaxy-mass density relation using the hydrodynamical simulation and 
  explore the physical origin of the scale dependent biasing on small 
  scales.

  The purpose of this paper is to investigate the galaxy-mass density 
  relation and its time evolution analytically. 
  Using the Edgeworth expansion, we will derive a general formula for 
  the conditional mean. Our main result is equation (\ref{dg-dm3}). 
  The final expression can be written in terms of the joint moment 
  of the galaxy and the total mass. Since the gravity induces the 
  time evolution 
  of the joint moments, the galaxy-mass density relation cannot be static. 
  Neglecting the merging and galaxy formation process, 
  we calculate the joint 
  moments using the perturbation theory of the Newtonian cosmology. Then we 
  study the time evolution of the galaxy-mass density relation. 
  We find that the estimation of higher order moments such as 
  skewness using the conditional mean could be affected by the 
  stochasticity. 

We organize this paper as follows. In section \ref{sec: formalism},  
the basic description of the stochastic biasing is explained. Within the 
formalism, a general formula for the galaxy-mass density relation is derived. 
Then we examine the time evolution of the galaxy-mass density relation 
in section \ref{sec: evolution}. 
The final section is devoted to the summary and the discussions.    
%
%
%
\section{Stochastic biasing and galaxy-mass density relation}
\label{sec: formalism}
%
%
%
%
%
%
\subsection{Basic description}
\label{sec: formalism1}
%
%
%
Let us recall the definitions of $\dm$ and $\dg$: 
  \begin{equation}
\dm(x)\equiv\frac{\rho(x)-\bar{\rho}}{\bar{\rho}},~~~~~     
\dg(x)\equiv\frac{n_g(x)-\bar{n}_g}{\bar{n}_g}, 
  \end{equation}
where variables with overbars indicate the homogeneous averaged density. 
In this paper, we treat the smoothed density fields $\dm(R)$ and 
$\dg(R)$ using the spherical top-hat window function $W_R(x)$ with the 
smoothing radius $R$: 

\begin{equation}
  \delta_{m,g}(R)\equiv \int d^3x~W_R(x)\delta_{m,g}(x). 
\end{equation}
Importantly, the one-point distribution function itself depends on the smoothing 
scale. As the smoothing scale becomes larger, the distribution approaches to 
the Gaussian distribution.   

The statistical information of the galaxy biasing is obtained from 
the joint probability distribution function (PDF) of the galaxy and 
the total mass distribution. In our formulation, 
$\dm$ and $\dg$ are regarded as the independent stochastic variables. 
To represent their stochasticity, we introduce the auxiliary random fields 
 $\Dm(x)$ and $\Dg(x)$. On large scales, the smoothed fields $\delta_{m,g}(R)$ 
could be regarded as nearly Gaussian fields. The density fields can be 
expressed as functions of the Gaussian variables and expanded in powers of 
these variables whose variances are small enough. Denoting such Gaussian 
variables as $\Delta_{m,g}$, the distributions $\dm$ and $\dg$ are 
given by 
\begin{equation}
  \dm=f(\Dm, \Dg),~~~~~\dg=g(\Dm, \Dg). 
  \label{initial-con}
\end{equation}
Here, we wrote equation (\ref{initial-con}) in somewhat schematical 
way. The "functions" $f$ and $g$ are not assumed as the local functions. 
The generating function for 
the one-point functions $\dm(R)$ and $\dg(R)$ becomes 
(\cite{taruya99}) 
\begin{eqnarray}
{\cal Z}(J_m,J_g)&\equiv&
\left\la e^{i\left\{J_m\dm(R)+J_g\dg(R)\right\}}\right\ra
\nonumber\\
&=&\int\int{\cal D}\Delta_m(x){\cal D}\Delta_g(x)~
{\cal N}^{-1}\exp\left[-(\Delta_m,\Delta_g)\mbox{\boldmath$G$}^{-1}
\left(
\begin{array}{c}
\Delta_m \\ \Delta_g
\end{array}
\right)\right]
\nonumber 
\\
&&~~~~~~~~\times
\exp{\left[i\int d^3x W_R(x)\left\{J_m\delta_m(\Delta_{m,g})+J_g
      \delta_g(\Delta_{m,g})\right\}\right]},
\label{Z}
\end{eqnarray}
where ${\cal N}$ is a normalization constant and 
$\mbox{\boldmath$G$}$ is a $2\times2$ matrix which gives the stochastic 
property of $\Dm$ and $\Dg$. The joint moments of $\dm$ and $\dg$ 
are deduced from the generating function as 
\begin{equation}
  \la[\dm(R)]^j[\dg(R)]^k\ra=i^{-(j+k)}
  \left.\frac{\partial^{(j+k)}{\cal Z}}{\partial J_m^j \partial J_g^k}\right|_
  {J_m=J_g=0}.
  \label{joint}
\end{equation}
The ensemble average $\la\cdots\ra$ is taken with respect to $\Dm$ 
and $\Dg$. Once we obtain the joint moments (\ref{joint}), the joint 
PDF ${\cal P}(\dm,\dg)$ for the 
smoothed density fields $\dm(R)$ and $\dg(R)$ is constructed: 
\begin{equation}
  {\cal P}(\dm,\dg)=\frac{1}{(2\pi)^2}\int\int dJ_m dJ_g
~{\cal Z}(J_m,J_g)~e^{-i(J_m\dm+J_g\dg)}, 
\label{joint-PDF}
\end{equation}
where we simply denotes $\delta_{m,g}(R)$ as $\delta_{m,g}$. 

We should keep in mind that the gravitational instability usually 
affects the galaxy and the total mass distribution. Accordingly, 
the functions $f$ and $g$ in the assumption (\ref{initial-con}) 
are determined by the gravitational dynamics and they would be 
generally non-local and time dependent (see section \ref{sec: evolution}). 
Since the resulting joint PDF also depends on time, we would 
have the time evolving biasing relation. 
%
%
%
%
%
%
%
%
\subsection{Stochastic biasing in the weakly 
non-linear regime: conditional mean and biasing scatter}
\label{subsec: conditional}
%
%
%
To proceed further, we shall employ the 
perturbative approach, hereafter. Taking the variables $\Delta_{m,g}$ 
as seeds of perturbation, $\dg$ and $\dm$ can be expressed as 
\begin{equation}
\delta_{m,g}(\Dm, \Dg)=\delta_{m,g}^{(1)}+\delta_{m,g}^{(2)}+\cdots,   
\label{series}
\end{equation}
where the $n$-th order perturbed quantities $\delta_{m,g}^{(n)}$ 
are of the same order as $[\Delta_{m,g}]^n$. 
Substituting the expansion 
(\ref{series}) into the joint moment (\ref{joint}), 
the non-vanishing lowest order quantities become the second moments.
They are characterized by the three parameters:  
\begin{equation}
  \sigm^2\equiv\la[\dm^{(1)}]^2\ra,~~~~~ 
  b^2\equiv\frac{\la[\dg^{(1)}]^2\ra}{\la[\dm^{(1)}]^2\ra},~~~~~
  r\equiv\frac{\la\dg^{(1)}\dm^{(1)}\ra}{\sqrt{\la[\dg^{(1)}]^2\ra
      \la[\dm^{(1)}]^2\ra}}.   
\label{tree-para}
\end{equation}
$\sigm$ is regarded as the variance of the total mass, $b$ is the 
linear biasing parameter and $r$ is the cross correlation. Note that 
the parameter $r$ reflects the stochastic property of 
$\dm$ and $\dg$. At the lowest order level, 
the deterministic biasing relation (\ref{linear-bias}) 
holds if $r=1$. 
 
The higher order perturbations leads to the non-linear galaxy biasing 
which describes the relation between the non-Gaussian distributions of 
galaxy and the total mass. Note that 
the higher order correction also affects the stochastic property of 
$\dm$ and $\dg$. To investigate the galaxy biasing in the weakly 
non-linear regime, we should analyze the non-linearity and the 
stochasticity separately.  

Several authors proposed to use the conditional mean of 
the galaxy distribution defined by (\cite{DL98}, \cite{cenII}) 
\begin{equation}
  \la\dg\ra|_{\dm}=\int d\dg ~\dg {\cal P}(\dg|\dm)~~~; ~~~
  {\cal P}(\dg|\dm)=\frac{{\cal P}(\dg,\dm)}{{\cal P}(\dm)},
  \label{conditional-mean}
\end{equation}
where ${\cal P}(\dg|\dm)$ denotes the conditional PDF at a 
given $\dm$. From the conditional mean $\la\dg\ra|_{\dm}$,  
we can know the non-linear relation of the biasing as the function 
of $\dm$.  
We should keep in mind that the conditional PDF has a width around the 
conditional mean $\cmg$. The {\it biasing scatter} can be introduced 
to measure the stochasticity in the biasing relation (\cite{DL98}):
\begin{equation}
  \label{scatter}
  \eps=\dg-\la\dg\ra|_{\dm}.
\end{equation}
The stochastic property of the non-linear biasing 
relation is understood from the variance and the higher moments of 
$\eps$. 
 
Consider the density fields on large scales.  
Owing to the assumption (\ref{initial-con}) and the 
perturbative treatment (\ref{series}), 
we can obtain the analytic expression for the conditional mean 
$\la\dg\ra_{\dm}$. The non-linear galaxy-mass density relation 
using the Edgeworth expansion leads to the expression 
\begin{equation}
   \la\dg\ra|_{\dm}=c_1\dm+\frac{c_2}{2}(\dm^2-\la\dm^2\ra)+\cdots.
\label{stochastic-relation} 
\end{equation}
We will show in the next subsection that the coefficients $c_1$ and 
$c_2$ are expressed in terms of the joint moments of $\dm$ and 
$\dg$ (see eq.[\ref{coeff-gal}]). 

Before deriving the galaxy-mass density relation, 
we mention the differences between the relation 
(\ref{stochastic-relation}) and (\ref{nl-bias}).  
From the relation (\ref{nl-bias}), Fry \& Gazta\~naga (1993) 
derived the useful formulae which relate the higher order 
moments of the galaxies with that of the total mass. For the 
third moments, we have 
\begin{equation}
  S_{3,g}=\frac{1}{b}(S_{3,m}+3\frac{b_2}{b}), 
\label{tree-bias-para} 
\end{equation}
where the linear biasing parameter $b_1$ is simply denoted as $b$. 
The quantities $S_{3,m}$ and $S_{3,g}$ are the 
skewness of the total mass and the galaxy distribution, respectively. 
In the tree level perturbation theory, these can be written as 
\begin{equation}
  S_{3,m}=3\frac{\la[\dm^{(1)}]^2\dm^{(2)}\ra}{\la[\dm^{(1)}]^2\ra},~~~~~
  S_{3,g}=3\frac{\la[\dg^{(1)}]^2\dg^{(2)}\ra}{\la[\dg^{(1)}]^2\ra}.
\label{skewness}
\end{equation}
%
For the galaxy-mass density relation 
(\ref{stochastic-relation}), we can also obtain the equation 
similar to (\ref{tree-bias-para}) (see eqs.[\ref{coeff-gal}] 
and [\ref{coeff-mass}]). 
We will see in section \ref{subsec: joint-moments}  
that the formula (\ref{tree-bias-para}) is recovered 
when the cross correlation $r$ becomes unity. Hence, the 
galaxy-mass density relation (\ref{stochastic-relation}) can be 
regarded as a generalization of the relation (\ref{nl-bias}) 
examined by Fry \& Gazta\~naga (1993).
%
%
%
%
\subsection{Derivation of galaxy-mass density relation: a general formula}
\label{subsec: derivation}
%
%
We are in a position to derive the galaxy-mass density relation 
(\ref{stochastic-relation}) as a conditional mean. 
In section \ref{sec: formalism1}, we defined 
the generating function 
${\cal Z}(J_m,J_g)$ to construct the joint PDF ${\cal P}(\dm,\dg)$. 
Using ${\cal Z}$, we obtain the generating function for the 
connected diagrams:
\begin{eqnarray}
&&  {\cal W}(J_m,J_g)\equiv \ln{\cal Z}
    = \sum_{j,k=0} \frac{\lambda_{jk}}{j!~k!}~
    \left(iJ_m \sqrt{\varm}\right)^j
    \left(iJ_g \sqrt{\varg}\right)^k, 
  \nonumber
\end{eqnarray}
where we define
\begin{eqnarray}
&&  \lambda_{jk}=\frac{\la(\dm)^j(\dg)^k\ra_c}
{\la(\dm)^2\ra^{j/2}~\la(\dg)^2\ra^{k/2}}.
\nonumber
\end{eqnarray}
$\la\cdots\ra_c$ denotes the connected part of the moments. 
By definition, we have $\lm_{02}=\lm_{20}=1$. 
The perturbative treatment (\ref{series}) 
implies that $\lambda_{ij}$ satisfies the following scaling relation. 
Using the variable $\sigm$ defined by equation (\ref{tree-para}) and   
assuming that the lowest order variance of the galaxies 
$\la[\dg^{(1)}]^2\ra$ is of the same order of 
magnitude as that of the total mass, we have 
\begin{equation}
  \lambda_{jk}=S_{jk}\sigm^{j+k-2}+{\cal O}(\sigm^{j+k}).
\label{scaling-relation}
\end{equation}
The variable $S_{ij}$ is of the order of unity and we obtain 
$S_{20}=S_{02}=1$. 

As long as the variance $\sigm$ is small, treating $\sigm$ 
as an expansion parameter, we can expand the 
generating function ${\cal W}$. Thus we proceed to 
approximate the joint PDF ${\cal P}(\dg,\dm)$. The expansion 
of ${\cal W}$ is referred to as the Edgeworth expansion (\cite{JWACB95}, 
\cite{BK95},\cite{chodorowski97}). 
Let us write the conditional mean $\cmg$ as 
\begin{equation}
\la\dg\ra|_{\dm}=\frac{{\cal A}}{{\cal P}(\dm)}~~;~~~
{\cal A}=\int d\dg~\dg{\cal P}(\dm,~\dg).
\label{dg-dm}
\end{equation}
Now, define the normalized variables:
\begin{equation}
\mu=J_m\sqrt{\varm},~~~~
\nu=J_g\sqrt{\varg},~~~~
x=\frac{\dm}{\sqrt{\varm}},~~~~
y=\frac{\dg}{\sqrt{\varg}}.
\label{new-var}
\end{equation}
Using the generating function ${\cal W}$, the numerator and the denominator 
of equation (\ref{dg-dm}) are expressed in terms of the above variables:
\begin{equation}
{\cal A}= \frac{1}{(2\pi)^2}
\left(\frac{\varg}{\varm}\right)^{1/2}
\int dy\int \int d\mu d\nu ~y e^{-i(\mu x+\nu y)+{\cal W(\mu,\nu)}}, 
\label{A}
\end{equation}
\begin{equation}
{\cal P}(\dm)= \frac{1}{2\pi}
\left(\frac{1}{\varm}\right)^{1/2}
\int d\mu e^{-i\mu x+{\cal W}(\mu,0)}.
\label{PDF-dm}
\end{equation}
Applying the Edgeworth expansion to the above equations, 
we can obtain the galaxy-mass density relation. 
The details of the calculation are described in Appendix A. 
Assuming that the normalized variables $x$ and $y$ are of the order of unity 
and using the 
scaling relation (\ref{scaling-relation}), we obtain the final results of 
the expansion up to ${\cal O}(\sigm^2)$ (\cite{chodorowski97},
\cite{BK95}):
\begin{eqnarray}
&&  {\cal A}=\frac{e^{-x^2/2}}{\sqrt{2\pi}}
\left(\frac{\varg}{\varm}\right)^{1/2}\left[
\lm_{11}H_1(x)+\left\{\frac{\lm_{21}}{2}H_2(x)+\frac{\lm_{30}}{6}H_4(x)
\right\}\right.
\nonumber \\
&&~~~~~~~~~~~\left.
+\left\{\frac{\lm_{31}}{6}H_3(x)+
\left(\frac{\lm_{11}\lm_{40}}{24}+\frac{\lm_{30}\lm_{21}}{12}\right)
H_5(x)+\frac{\lm_{11}\lm_{30}^2}{72}H_7(x)\right\}\right], 
\label{reduced-A}
\end{eqnarray}
and 
\begin{equation}
{\cal P}(\dm)=\frac{e^{-x^2/2}}{\sqrt{2\pi}\sqrt{\varm}}
\left[1+\frac{\lm_{30}}{6}H_3(x)+
\left\{\frac{\lm_{40}}{24}H_4(x)+\frac{\lm_{30}^2}{72}H_6(x)\right\}\right],
  \label{dm-PDF}
\end{equation}
where $H_n(x)$ is the Hermite polynomial defined by 
\begin{equation}
  H_n(x)= e^{x^2/2}\left(-\frac{d}{dx}\right)^n e^{-x^2/2}.
\label{Hermite}
\end{equation}
%
Substituting the results (\ref{reduced-A}) and (\ref{dm-PDF}) into 
equation (\ref{dg-dm}), the conditional mean $\cmg$ in the weakly 
non-linear regime becomes
\begin{equation}
  \label{dg-dm2}
  \la\dg\ra|_{\dm}=\sqrt{\varg}~\left[
\lm_{11}x+{\cal F}_1(x)+{\cal F}_2(x)\right].
\end{equation}
where ${\cal F}_n(x)$ denotes the terms of the order 
${\cal O}(\sigm^n)$ given by
\begin{eqnarray}
{\cal F}_1(x)&=&\frac{\lm_{21}}{2}H_2(x)+
\frac{\lm_{11}\lm_{30}}{6}\left[H_4(x)-xH_3(x)\right],
\nonumber \\
{\cal F}_2(x)&=&\frac{\lm_{31}}{2}H_3(x)+
\frac{\lm_{30}\lm_{21}}{12}\left[H_5(x)-H_2(x)H_3(x)\right]+
\frac{\lm_{11}\lm_{40}}{24}\left[H_5(x)-xH_4(x)\right]
\nonumber \\
&&~~~+\frac{\lm_{11}\lm_{30}^2}{72}
\left[H_7(x)-xH_6(x)-2H_3(x)H_4(x)+2xH_3(x)^2\right].
\nonumber
\end{eqnarray}
We should keep in mind that $\lm_{ij}$ still contains the terms higher than 
${\cal O}(\sigm^{i+j})$ and we do not approximate the variance of 
galaxy distribution $\varg$ in deriving the expression (\ref{dg-dm2}). 

We first consider the conditional mean to the order 
${\cal O}(\sigm)$. Since we can drop the higher order 
correction to $\lm_{ij}$ with $i+j\geq3$, we simply replace 
$\lm_{ij}$ with $S_{ij}\sigm^{i+j-2}$. 
As for the variance and the covariance of $\dm$ and $\dg$, 
the definition (\ref{tree-para}) gives
\begin{eqnarray}
&&  \varm=\sigm^2,~~~~\varg=b^2\sigm^2,~~~~~\lm_{11}=r  
\nonumber
\end{eqnarray}
This is correct within the tree-level analysis. 
Then the weakly non-linear relation (\ref{dg-dm2}) up to 
${\cal O}(\sigm)$ is written in terms of the variable $\dm$ 
as 
\begin{equation}
  \la\dg\ra|_{\dm}= c_1 ~\dm +\frac{c_2}{2}
\left(\dm^2 - \la\dm^2\ra \right)+
{\cal O}(\sigm^2).
\label{dg-dm3} 
\end{equation}
The coefficients $c_1$ and $c_2$ becomes
\begin{equation}
c_1=b~r,~~~~~~~c_2=b~(S_{21}-rS_{30}). 
\label{coeff-gal}
\end{equation}
%
The expressions (\ref{dg-dm3}) and (\ref{coeff-gal}) are our main results.

To understand the stochastic property in the biasing relation, 
the mass density-galaxy relation should be noted: 
\begin{eqnarray}
  \la\dm\ra|_{\dg}&=&\int d\dm ~\dm {\cal P}(\dm|\dg), 
\nonumber \\
&=&\frac{{\cal B}}{{\cal P}(\dg)}~~;~~~
{\cal B}=\int d\dm~\dm{\cal P}(\dm,\dg). 
\nonumber
\end{eqnarray}
Derivation of non-linear mass density-galaxy relation is similar to 
that of $\cmg$. For the expression (\ref{dg-dm2}), 
we only have to replace the variables $\varg$, $\lm_{ij}$ and $x$ 
with $\varm$, $\lm_{ji}$ and $y$, respectively. We here 
write down the result in terms of $\dg$ up to ${\cal O}(\sigm)$: 
\begin{equation}
  \la\dm\ra|_{\dg}= d_1 ~\dg +\frac{d_2}{2} 
  \left(\dg^2 - \la\dg^2\ra \right)+
{\cal O}(\sigm^2). 
\label{dm-dg2} 
\end{equation}
Then the coefficients $d_1$ and $d_2$ become
\begin{equation}
d_1=\frac{r}{b},~~~~~~~d_2=\frac{1}{b^2}(S_{12}-rS_{03}). 
\label{coeff-mass}
\end{equation}
We observe that the conditional mean 
(\ref{dm-dg2}) does not coincide with the result (\ref{dg-dm3}) due to  
the presence of the cross correlation $r$ and the fact that 
$S_{ij}$ is not symmetric. It will be shown in the next section 
that both relations give the same relation when the 
cross correlation $r$ becomes unity. 

%
We next clarify the effect of 
the higher order corrections on the galaxy-mass relation $\cmg$. 
When we take into account the ${\cal F}_2(x)$ term, the 
${\cal O}(\sigm^2)$ contribution requires the third order perturbations.  
In this case, we cannot ignore the ${\cal O}(\sigm^2)$ 
correction for $\lm_{ij}$ 
with $i+j=2$. Then the evaluation of the first term in 
the right hand side of (\ref{dg-dm2}) is modified. Using the definition 
(\ref{tree-para}), we can write
\begin{equation}
\sqrt{\varg}~\lm_{11}x=\frac{\la\dg\dm\ra}{\la\dm^2\ra} \dm=
\left[br+{\cal O}(\sigm^2)\right]\dm, 
\label{renorm-term}
\end{equation}
where the order ${\cal O}(\sigm^2)$ terms come from the loop 
corrections: 
\begin{eqnarray}
&&  \sigm^{-2}\left[
\la\dm^{(2)}\dg^{(2)}\ra+\la\dm^{(3)}\dg^{(1)}\ra+
\la\dm^{(1)}\dg^{(3)}\ra
-br\left(\la[\dm^{(2)}]^2\ra+2\la\dm^{(1)}\dm^{(3)}\ra\right)\right].
\nonumber
\end{eqnarray}
Hence we redefine equation (\ref{renorm-term}) as $b_{ren}r_{ren}\dm$ 
to the ${\cal O}(\sigm^2)$ terms. 
The parameters $b_{ren}$ and $r_{ren}$ are the renormalized 
quantities for the biasing parameter $b$ and the cross correlation $r$ 
given by the tree-level perturbation. 
For $\lm_{ij}$ with $i+j\geq3$, the 
correction terms are higher than ${\cal O}(\sigm^3)$, 
which can be verified from the scaling law (\ref{scaling-relation}).  
Rewriting the expression (\ref{dg-dm2}) in terms 
of $\dm$ by using the definition (\ref{Hermite}), 
we get the non-linear galaxy-mass density 
relation in accuracy of ${\cal O}(\sigm^2)$: 
\begin{equation}
  \la\dg\ra|_{\dm}= \tilde{c}_1 ~\dm +\tilde{c}_2 
  \left(\dm^2 - \la\dm^2\ra \right)+
  \tilde{c}_3~\dm^3+{\cal O}(\sigm^3), 
  \label{dg-dm4}
\end{equation}
where the coefficients $\tilde{c}_i$ are
\begin{eqnarray}
  \tilde{c}_1&=&b_{ren}r_{ren}+
  \sigm^2\left[S_{30}S_{21}-rS_{30}^2-\frac{1}{2}(S_{31}-rS_{40})
  \right], 
\nonumber \\
  \tilde{c}_2&=&\frac{b}{2}(S_{21}-rS_{30}), 
\nonumber \\
  \tilde{c}_3&=&\frac{b}{6}(S_{31}-rS_{40})-
\frac{b}{2}(S_{30}S_{21}-rS_{30}^2).
\nonumber 
\end{eqnarray}
%
We immediately see the 
difference between the above result (\ref{dg-dm4}) and 
the tree-level result (\ref{dg-dm3}). In addition to 
the contribution of $\dm^3$, the coefficient $\tilde{c}_1$ has the 
extra terms except for the renormalized parameters, which are absent 
from the coefficient $c_1$. The result says that the linear biasing relation 
in the weakly non-linear regime is different from the linear theory 
prediction even for $\dm\ll1$. 
Our results are in good agreement with the 
analyses of the two point correlation function for the galaxy distribution 
(\cite{SW97}).
%
%
%
\section{Time evolution of galaxy-mass density relation}
\label{sec: evolution}
%
%
%
%
The previous section has been devoted to the discussion of the 
galaxy-mass density relation. 
In this section, we investigate 
the time evolution of it. 
The evolution 
equations and the initial conditions are given in section 
\ref{sec: formalism2}. Within this model,  
we calculate the time-dependence of the 
joint moments $S_{ij}$ explicitly and the relationship between the conditional 
means $\cmg$ and $\cmm$ is clarified in Sec\ref{subsec: joint-moments}. 
In section \ref{subsec: evolution-cmg}, 
we will study the time evolution of the conditional mean 
$\cmg$ in order to reveal the influence of initial stochasticity. 
%
%
%
%
%
\subsection{Evolution equations and initial conditions}
\label{sec: formalism2}
%
%
%
%
%
%
The time evolution of the galaxy biasing 
induced by the gravity has been studied by several authors 
(\cite{F96}, \cite{TP98}, \cite{taruya99}). In these articles, 
the density fluctuation $\dm$ is assumed to be evolved  
following the equation of continuity and the 
Euler equation. On large scales, 
the total mass distribution moves along the irrotational velocity flow. 
It is convenient to define the velocity divergence 
$\theta\equiv\nabla\cdot\v/(aH)$, 
where $a$ is the scale factor of the universe and $H$ is the Hubble 
parameter. For the galaxy distribution, we consider the epoch after 
galaxy formation 
and assume that the merging process is negligible. 
The distribution $\dg$ should satisfy the equation of 
the continuity whose velocity field is determined by 
the gravitational potential. 
Then the evolution equations for the total mass and the galaxies become 
\begin{eqnarray}
&& \frac{\partial\delta_m}{\partial t}+H\theta+
\frac{1}{a}\nabla\cdot(\delta_m\v)=0,
\label{basic-eq1}
\\
&&  \frac{\partial\theta}{\partial t}+\left(1-\frac{\Omega}{2}+
\frac{\Lambda}{3H^2}\right)H\theta+\frac{3}{2}H\Omega\delta_m
+\frac{1}{a^2H}\nabla\cdot(\v\cdot\nabla)\v=0,
\label{basic-eq2}
\end{eqnarray}
and 
\begin{equation}
  \frac{\partial\delta_g}{\partial t}+H\theta+
\frac{1}{a}\nabla\cdot(\delta_g\v)=0.
\label{basic-eq-gal}
\end{equation}
The variable $\Lambda$ is the cosmological constant and $\Omega$ is the 
density parameter defined by 
\begin{equation}
\Omega\equiv \frac{8\pi G}{3}\frac{\bar{\rho}}{H^2}.  
\end{equation}
%
Equations (\ref{basic-eq1}), (\ref{basic-eq2}) and  (\ref{basic-eq-gal}) are 
our basic equations for the time evolution of $\dm$ and $\dg$. 
The same situation has been studied by Fry (1996) in the case of the 
deterministic biasing (\ref{nl-bias}). 

Next, we consider the initial condition given at an 
 initial time $t_i$. 
The total mass fluctuation $\dm$ is produced during the very early 
stage of the universe, whose initial distribution may have 
the random Gaussian statistics. 
We regard such fluctuation as $\Dm(x)$. The 
gravitational instability induces the deviation from the Gaussian 
statistics and the galaxy formation does not affect the 
evolution of $\dm$ on large scales. 
We give the initial condition $\dm=f(\Dm)$ from the perturbative 
solutions by dropping the decaying mode, which leads to  
the non-local form of the function $f(\Dm)$ (see Appendix B).

On the other hand, the fluctuation of the galaxy number density 
is induced by the galaxy formation. 
To give the initial non-Gaussian distribution $\dg=g(\Dg)$, 
we need to know the galaxy formation processes. Currently, it is not 
formidable. Here, we treat $g(\Dg)$ as a 
parameterized function, 
whose unknown parameters are determined by the observation of galaxies. 
Assuming $g(\Dg)$ as a local function of $\Dg$, we take 
\begin{equation}
  g(\Dg)=\Dg+\frac{h}{6}(\Dg^2-\langle\Dg^2\rangle)+\cdots.
\label{init-g}
\end{equation}
%

Due to the assumptions stated above, 
in the perturbative regime, the relation between the galaxy and 
the total mass distribution can be characterized completely if the 
stochastic property of the 
  Gaussian variables $\Dm$ and $\Dg$ is given. It is 
expressed in terms of the 
  three parameters : 
\begin{equation}
  \sigma_0^2={\langle\Delta_m^2\rangle},~~~~
  b_0^2=\frac{\langle\Delta_g^2\rangle}{\langle\Delta_m^2\rangle},~~~~
  r_0=\frac{\langle\Delta_g\Delta_m\rangle}
{\left(\langle\Delta_m^2\rangle\langle\Delta_g^2\rangle\right)^{1/2}},
\label{stochastic-para}
\end{equation}      
which is equivalent to giving the matrix $\mbox{\boldmath$G$}$ in the 
generating function (\ref{Z}). 
The variables 
 $b_0$, $r_0$ and $\sigma_0$ corresponds to the initial biasing parameter, 
the initial cross correlation and the initial variance of the total mass, respectively. 
The variance $\sigma_0$ is related to the initial power spectrum $P(k)$ as 
\begin{eqnarray}
  &&\sigma_0^2(R)=\int\frac{d^3\k}{(2\pi)^3}\tilde{W}_R^2(kR)P(k) 
~~~;~~~\tilde{W}(kR) =\frac{2}{(kR)^3}[\sin{(kR)}-kR\cos{(kR)}], 
\label{init-var}
\end{eqnarray}
where $\tilde{W}(kR)$ is the top-hat 
window function  in the Fourier space. Hereafter, we simply assume that 
the parameters $b_0$ and $r_0$ are constant, which is consistent with 
the fact that there is no evidence of the scale-dependent biasing on 
large scales (\cite{Man98}). 
%
%
%
%
%
\subsection{Variance, covariance and joint moments}
\label{subsec: joint-moments}
%
%
%
%
Within the above prescription,  
we will investigate the time evolution of the galaxy-mass density relation 
in the weakly non-linear regime.  
In the lowest order of the perturbation, 
the conditional means (\ref{dg-dm3}) and (\ref{dm-dg2}) are 
written in terms of the variance of the total mass $\sigm$, 
the biasing parameter $b$, the cross correlation 
$r$ and the joint moments $S_{ij}$ with $i+j=3$.  

The parameters $\sigm$, $b$ and $r$ are obtained from the linear order 
solutions of evolution equations (\ref{basic-eq1}), (\ref{basic-eq2}) and 
(\ref{basic-eq-gal}). The solutions satisfying the initial  
conditions in the previous subsection become 
\begin{eqnarray}
\delta_m^{(1)}(x,t)&=&\Delta_m(x)~D(t),
\label{m-sol-1st}
 \\
\delta_g^{(1)}(x,t)&=&\Delta_m(x)~(D(t)-1)+\Delta_g(x),
\label{g-sol-1st}
\end{eqnarray}
where the function $D(t)$ denotes the solution of growing mode 
by setting $D(t_i)=1$, which satisfies (\cite{Peebles}) 
\begin{eqnarray}
&&  \ddot{D}+2H\dot{D}-\frac{3}{2}H^2\Omega D=0. 
\label{grow-eq}
\end{eqnarray}
We have $D(t)=a(t)/a(t_i)$ in Einstein-de Sitter universe ($\Omega=1$). 
Substituting (\ref{m-sol-1st}) and (\ref{g-sol-1st}) into (\ref{tree-para}), 
the time dependent parameters are evaluated from the knowledge 
(\ref{Z}) and (\ref{stochastic-para}) as follows 
(\cite{F96}, \cite{taruya99}, \cite{TP98}):
\begin{eqnarray}
\sigm(t)&=&\sigma_0~D,
  \nonumber
\\
b(t)&=&\frac{\sqrt{(D-1)^2+2b_0r_0(D-1)+b_0^2}}{D}, 
  \label{linear-t-bias-parameter}
\\
r(t)&=&b^{-1}(t)\left(\frac{D-1+b_0r_0}{D}\right). 
  \nonumber
\end{eqnarray}
We see that the initial cross correlation $r_0=1$ leads to $r=1$. Thus, 
in our prescription of the time evolution,   
the stochasticity in the galaxy biasing comes from the initial 
conditions.

For the joint moments $S_{ij}$, we must 
calculate the second order perturbations by solving the evolution equations.  
In Appendix B, 
the solutions of second order perturbation are summarized and the 
computation of $S_{ij}$ is explained. The time 
evolution of the joint moments are parameterized by the initial conditions,   
$b_0$, $r_0$, $h$, and the power spectrum $P(k)$. The results in 
Appendix B are 
\begin{eqnarray}
  S_{30}&=&\frac{34}{7}-\gamma, 
  \label{s30} 
\\
  S_{21}&=&\frac{1}{3b(t)}
  \left[\left(\frac{34}{7}-\gamma\right)
    \left\{\frac{D^2-1}{D^2}+\frac{2(D-1+b_0r_0)}{D}\right\}\right.
\nonumber \\
&&+\left.\left(6-\gamma\right)\frac{(D-1)(b_0r_0-1)}{D^2}
+h~I_W\frac{(b_0r_0)^2}{D^2}\right],
  \label{s21} 
\\
  S_{12}&=&\frac{1}{3b^2(t)}
  \left[\left(\frac{34}{7}-\gamma\right)
    \left\{\frac{(D-1+b_0r_0)^2}{D^2}+\frac{2(D^2-1)(D-1+b_0r_0)}{D^3}
    \right\}\right.
\nonumber \\
&&+\left(6-\gamma\right)
\left\{\frac{2(D-1)^2(b_0r_0-1)}{D^3}
  +\frac{(D-1)((b_0r_0-1)^2+b_0^2-1)}{D^3}\right\}
\nonumber \\
&&+\left.h~I_W\frac{2b_0r_0(b_0r_0(D-1)+b_0^2)}{D^3}\right],
  \label{s12} 
\\
S_{03}&=&\frac{1}{b^3(t)}
  \left[\left(\frac{34}{7}-\gamma\right)
    \frac{(D^2-1)(D-1+b_0r_0)^2}{D^4}\right.
\nonumber \\
&&+\left(6-\gamma\right)
\frac{(D-1)(D-1+b_0r_0)\left\{(b_0r_0-1)D+b_0^2+1-2b_0r_0\right\}}{D^4}
\nonumber \\
&&+\left.h~I_W\frac{\left\{b_0r_0(D-1)+b_0^2\right\}^2}{D^4}\right],
  \label{s03} 
\end{eqnarray}
where the numerical value $I_W$, which comes from the 
non-Gaussian initial distribution 
of the galaxies, is defined by 
\begin{eqnarray}
  && I_W=\sigma_0^{-4}\int\frac{d^3\k_1d^3\k_2}{(2\pi)^6}
\tilde{W}_R(|\k_1+\k_2|R)\tilde{W}_R(k_1R)\tilde{W}(k_2R)P(k_1)P(k_2). 
\nonumber
\end{eqnarray}
It has been checked by the Monte Carlo integration that $I_W$ is 
almost equal to unity. Hence, we can regard $h$ as the initial skewness 
of the galaxy distribution. The variable $\gamma$ is 
given by  
\begin{eqnarray}
&& \gamma=-\frac{d}{d(\log{R})}\cdot [\log{\sigma^2_0(R)}].
\nonumber
\end{eqnarray}
For the power spectrum with the single power-law $P(k)\propto k^n$, 
we have $\gamma=n+3$. 

By definitions (\ref{tree-para}) and (\ref{scaling-relation}), 
the moment $S_{30}$ is identical with the skewness of the total mass 
$S_{3,m}$ (see [\ref{skewness}]). The skewness of the galaxy 
distribution $S_{3,g}$ is related to the moment $S_{03}$ as 
$S_{3,g}=S_{03}/b(t)$. In the case of $r_0=1$, using these facts, 
and equations (\ref{s30})-(\ref{s03}), we can get the relation  
\begin{eqnarray}
&&  c_2=b(S_{21}-rS_{30})=\frac{b}{3}(b S_{3,g}-S_{3,m}).
\nonumber
\end{eqnarray}
Thus, in the deterministic case, the coefficients $c_1$ and $c_2$ 
can coincide with $b$ and $b_2$ given by equations 
(\ref{nl-bias}) and (\ref{stochastic-relation}). 
It should be 
recognized that the galaxy-mass density relation defined by equation  
(\ref{nl-bias}) does not coincide with the conditional 
mean $\cmg$ in general. 

Using the results (\ref{s30})-(\ref{s03}), 
we have another important conclusion for the galaxy-mass density 
relation. We can obtain 
\begin{eqnarray}
&&    S_{21}-S_{30}=-(S_{12}-S_{03}), 
\nonumber
\end{eqnarray}
for $r_0=1$. Taking $\dg=b\dm$, this means that the 
approximation of 
$\dm$ by the perturbative inversion of $\cmg$ coincides with 
the conditional mean $\cmm$ (eq.[\ref{dm-dg2}]). However, 
if we take into account the higher order corrections, 
we can expect that the conditional means $\cmg$ and $\cmm$ 
does not become equivalent even in $r_0=1$ case, because of the 
higher order contribution to the linear order coefficient of 
the conditional means (see eq.[\ref{dg-dm4}] below). 
%
%
%
%
%
%
%
\subsection{Evolution of $\cmg$}
\label{subsec: evolution-cmg}
%
%
%
%
Now consider the time evolution of the conditional mean 
$\cmg$ to see the influence of the initial stochasticity on the 
galaxy-mass density relation. For brevity, we only describe the analysis 
in the Einstein-de Sitter universe ($\Omega_0=1$). 

Using the results (\ref{linear-t-bias-parameter}), (\ref{s30}) 
and (\ref{s21}), the time evolution of the 
coefficients $c_1$ and $c_2$ can be examined. 
In Fig.1, we plot the coefficients $c_1$ and $c_2$ 
given by equation (\ref{coeff-gal}) as a function of the expansion factor $a(t)$. 
As the initial parameters, we chose $b_0=2.0$, $r_0=0.8$ and 
$h=3.0$ at $a(t_i)=1$ in Fig.1a. The solid line shows the coefficient $c_1$. 
The long-dashed, the short-dashed and the dotted lines denote the 
time evolution of the coefficient $c_2$ with the 
spectral index $n=-2$, $-1.5$ and $-1$ for the single 
power-law $P(k)\propto k^n$, respectively. 
The parameters in Fig.1b is the same as Fig.1a, except for the initial cross 
correlation for which we take $r_0=0.1$. These 
figures show that the coefficients 
$c_1$ and $c_2$ approach to unity and zero respectively, 
independent of the initial parameters and the spectral index. 
This asymptotic behavior can be ascribed to the attractivity of the 
gravitational force (\cite{F96}, \cite{TP98}, \cite{taruya99}). 
Fig.1c shows the illustrative example with the initial parameters 
$b_0=4.63$, $r_0=0.2$ and $h=6.96$ at $a(t_i)=1$. We set the spectral index 
$n=-1.41$. If we identify the initial time $a=1$ with the 
redshift parameter $z=3$ assuming the Einstein-de Sitter universe, 
the skewness and the bi-spectrum at present time 
(which corresponds to $a=4$ in our case) provide the consistent results with 
the observation of the Lick catalog (\cite{taruya99}). 
Although the initial value of the coefficient $c_2$ 
is comparable to that of $c_1$ because of the large initial skewness $h$, it 
rapidly decreases and becomes negligible due to the small initial cross 
correlation $r_0$. As a demonstration, we evaluate the galaxy-mass density 
relation $\cmg$ in the $(\dg, \dm)$-plane with the same initial 
parameters as Fig.1c. The galaxy-mass relation is plotted in 
Fig.2 by choosing the initial variance of the total mass $\sigma_0=0.1$. Each 
line in Fig.2 represents the snapshot at $a=1$ 
({\it solid line}), $a=2$ ({\it long-dashed line}), 
$a=4$ ({\it short-dashed line}) and $a=8$ ({\it dotted line}), respectively. 

Fig.1 and Fig.2 say that the coefficient $c_2$ is 
usually smaller than the linear coefficient $c_1$. 
We can also confirm this fact for various initial parameters. 
This result means that the non-linearity in the 
conditional mean is negligible on large scales. 
Therefore the result indicates that the relation between the galaxy and 
the total mass distribution is well-approximated by the linear 
relation $\dg=br\dm$. If we apply this fact to the estimation of 
the higher order moments for galaxy distribution, 
we expect that the moments of 
the galaxy are simply related to those of the total 
mass multiplied by the factor inferred from the 
linear relation. That is, regarding $\cmg$ as $\dg$, we obtain
\begin{equation}
\la(\dg)^n\ra\simeq(br)^n\la(\dm)^n\ra.  
\label{linear-var}
\end{equation}
However, we should keep in mind that there exists the scatter in 
the galaxy-mass relation. In section \ref{subsec: conditional}, 
we have defined the biasing scatter $\eps$ (see eq.[\ref{scatter}]). 
In the tree-level analysis, the variance of the biasing scatter becomes 
\begin{equation}
  \frac{\la\eps^2\ra}{\sigm^2}=(1-r^2)+{\cal O}(\sigm^2).
  \label{var-scatter}
\end{equation}
The variance vanishes only when $r_0=1$. We can also calculate 
the third order moment. The result up to the second order perturbation is 
\begin{equation}
  \frac{\la\eps^3\ra}{\sigm^4}=b^3\left[S_{03}-r^3S_{30}-3
    r\left(S_{12}-r S_{21}\right)\right]+{\cal O}(\sigm^2),
  \label{skew-scatter}
\end{equation}
where we used $c_1=br$. Substituting the expressions 
(\ref{s30})-(\ref{s03}) and (\ref{linear-t-bias-parameter}) 
into equation (\ref{skew-scatter}), it is easily shown 
that the third order moment of the biasing scatter also vanishes in the 
deterministic case, $r_0=1$. From equations (\ref{var-scatter})  
and (\ref{skew-scatter}),  the non-Gaussian 
scatter is expected to affect the simple relation of the 
higher order moments (\ref{linear-var}). 

To see the influence of the biasing scatter on the evaluation of 
higher order moments, we examine the skewness of the galaxy 
distribution. In Fig.3, we plot the time evolution of $S_{3,g}$ 
as a function of the expansion factor. For each figure, the solid line 
is the correct skewness obtained from the joint moment $S_{03}$ divided 
by $b(t)$ and the dashed line represents the skewness $\tilde{S}_{3,g}$ 
deduced from the galaxy-mass density relation $\cmg$. 
$\tilde{S}_{3,g}$ can be evaluated by equating the conditional 
mean $\cmg$ with $\dg$. The non-linear relation 
between $\dm$ and $\dg$ (\ref{dg-dm3}) implies 
\begin{equation}
  \tilde{S}_{3,g}=\frac{1}{c_1}(S_{3,m}+3\frac{c_2}{c_1}),  
\label{tilde-skewness}
\end{equation}
which is valid within the tree-level approximation (\cite{FG93}). We 
have already obtained the skewness of the total mass 
$S_{3,m}=\frac{34}{7}-\gamma$ and found that $S_{3,m}$ is usually 
larger than the coefficient $3(c_2/c_1)$. Fig.3a and Fig.3b 
have the same initial parameters as Fig.1a and Fig.1b, respectively, 
except for the power spectrum specified as the index $n=-2$.  
As we know from the previous subsection, $\tilde{S}_{3,g}$ 
exactly coincides with the skewness $S_{3,g}$ only in the deterministic 
case, i.e, $r_0=1$. The figures show that $\tilde{S}_{3,g}$ differs 
from $S_{3,g}$ in the presence of stochasticity clearly. 
For the smaller initial cross correlation $r_0$, the deviation of 
$\tilde{S}_{3,g}$ from $S_{3,g}$ becomes more significant (Fig.3b). 
Because of the small contribution of 
the non-linear coefficient $c_2$, $\tilde{S}_{3,g}$ approaches to 
the skewness of the total mass $S_{3,m}=3.86$ more rapidly than 
$S_{3,g}$ (see eq.[\ref{tilde-skewness}]).  
For the same parameterization as used in Fig.1c, 
Fig.3c shows that $\tilde{S}_{3,g}$ at 
present ($a=4$) underestimates 
the correct value of the skewness 
for the galaxy distribution due to the rapid relaxation to $S_{3,m}$. 
At $a=4$, the skewness $S_{3,g}$ from Fig.3c becomes 4.55, while 
we have $\tilde{S}_{3,g}=3.50$. The difference cannot be neglected. 
Thus there exist the cases that the galaxy-mass density 
relation as a conditional mean leads to incorrect result. 
This feature is common to the stochastic biasing. 
%
%
%
%
%
\section{Conclusion}
\label{sec: conclusion}
%
%
%
%
In this paper, 
we have analytically studied the galaxy-mass density relation 
in the framework of the stochastic biasing. 
Under the assumptions (\ref{initial-con}) and (\ref{Z}), 
in the weakly non-linear regime, 
we derived a general formula for the galaxy-mass 
density relation as a conditional mean. The main result 
in our analysis is equations (\ref{dg-dm3}) and (\ref{coeff-gal}). 
We have seen that the higher order contribution to the weakly non-linear 
galaxy-mass density relation can shift the coefficient of the linear 
galaxy-mass density relation, in addition to the non-linear term 
proportional to $\dm^3$ (see eq.[\ref{dg-dm4}]). This agrees 
with the analyses of the two point correlation function (\cite{SW97}). 
Using the formula for the conditional mean, we have further 
investigated the time evolution of the galaxy-mass density relation. 
To develop the analysis, we have made the assumptions: 
(i) the galaxy formation and the 
merging process can be ignored; 
(ii) the initial distribution of galaxies 
is given by the local function (eq.[\ref{init-g}]), while 
the total mass has the non-local initial condition. 
Our conclusions can be summarized as follows: 
\begin{itemize}
\item  The conditional mean could be different from the 
  deterministic biasing relation (compare the relation [\ref{tree-bias-para}] 
  and [\ref{coeff-gal}]). In the presence of the scatter, the perturbative 
  inversion of $\cmg$ does not recover the conditional mean $\cmm$. 
\item  The time evolution of the conditional mean $\cmg$  
  shows that the non-linear term of the conditional means 
  usually becomes negligible. 
  This suggests that the galaxy-mass density relation could be 
  approximated by the linear relation. 
\item We have found  
  that the time evolution of the skewness deduced from 
  the conditional mean $\cmg$ differs from that of the correct 
  skewness $S_{3,g}$ in the case of the small cross correlation. 
  This indicates that the stochasticity 
  could have an important role in the estimation of the higher 
  order moment of galaxies using the conditional mean. 
\end{itemize}

The assumptions (i) and (ii) might not be valid 
in more realistic situation. The merging process might 
become important in a real universe and the galaxy formation 
would introduce the non-local initial condition.  
Nevertheless, we believe that our qualitative conclusions will 
not be altered even if these processes are taken into account. 
It would be interesting to incorporate these effects into 
the time evolution of the stochastic biasing. 
We will attack this issue in the future. 

The conditional mean $\cmg$ 
is an important quantity to construct the analytic biasing model 
which is determined by the clustering properties of the halos 
(\cite{MW96}, \cite{MJM}, \cite{Catelan}, \cite{Sheth}). 
In the presence of the stochasticity, we must treat the 
conditional mean carefully when comparing it with the 
observation of the galaxy statistics, and vice versa. 
That is, the linear and non-linear biasing parameter ($b,~b_2$)  
estimated from the observation of the skewness or kurtosis 
does not necessarily give the mean biasing relation. They might provide 
a signal for the stochastic biasing. 
In the previous paper (\cite{taruya99}), we found that the parameter 
$b_2$ deduced from the skewness can become negative in the presence 
of the scatter. The recent observation from the Southern Sky Redshift 
Survey shows that the non-linear biasing parameter $b_2$ 
estimated by the skewness may become negative for the biasing parameter 
$b>1$ (\cite{BCCMBS98}). This might give an observational 
evidence for the stochastic biasing as long as our prescription is correct. 

The stochasticity is problematic when we  
get the relation between the galaxies and the total mass from the 
observation. As discussed by Dekel \& Lahav (1998), the situation may become  
more complicated in the redshift space. 
This makes it difficult to determine the cosmological parameter from 
the observation of the velocity field such as POTENT 
(\cite{BD89}). 
For the analysis in the redshift space, Scoccimarro, Couchman \& Frieman (1998) 
studied the influence of the non-linear biasing on the bi-spectrum of the 
galaxies in the deterministic case. We must investigate how 
the stochasticity and the non-linearity affect the galaxy biasing 
in the redshift space. In particular, 
the higher order statistics such as the skewness and the bi-spectrum 
should be explored. Extension of our formalism to the redshift 
space is straightforward and the analysis is now going on (\cite{taruya00}). 

Alternative approach to understand the stochastic property of the galaxy 
biasing is to seek the physical origin of the stochastic biasing itself. 
Blanton {\it et al}. (1998) explored the hidden variable to 
reduce the stochasticity in the relation 
between galaxies and total mass using the hydrodynamical simulation.  
They found that the scatter around the conditional mean of the galaxies 
becomes small if the temperature dependence is taken into account. 
To combine this approach with ours is also interesting. 
%
%
%
%
\acknowledgments
This work is partially supported by Monbusho 
Grant-in-Aid for Scientific Research 10740118.
%
%
%
%
%
%
%
\newpage
\appendix
%
%
%
%
\section*{Appendix A: Calculation of Edgeworth series}
%
%
In this appendix, we explain the calculation for the derivation of 
the galaxy-mass density relation as a conditional mean in section 
\ref{subsec: derivation}. 

We start to write down the generating function ${\cal W}$ 
expanded in powers of $\sigm$. 
From the scaling relation (\ref{scaling-relation}), the 
generating function up to the order ${\cal O}(\sigm^2)$ is obtained 
in terms of the normalized variables (\ref{new-var}): 
\begin{eqnarray}
&&  {\cal W}(\mu,~\nu)=-\frac{1}{2}(\mu^2+2\lm_{11}\mu\nu+\nu^2)
\nonumber\\
&&~~+\frac{1}{6}\left\{
\lambda_{30}(i\mu)^3+3\lambda_{21}(i\mu)^2(i\nu)+3\lambda_{12}(i\mu)(i\nu)^2+
\lambda_{03}(i\nu)^3\right\}
\label{expand-W}\\
&&~~+\frac{1}{24}\left\{
\lambda_{40}(i\mu)^4+
4\lambda_{31}(i\mu)^3(i\nu)+
6\lambda_{22}(i\mu)^2(i\nu)^2+
4\lambda_{13}(i\mu)(i\nu)^3+
\lambda_{04}(i\nu)^4\right\}
\nonumber \\
&&~~+{\cal O}(\sigm^3).
\nonumber 
\end{eqnarray}
Note that the first line in the right hand side of (\ref{expand-W}) is of 
the order of unity. The second and the third lines are the 
terms of ${\cal O}(\sigm)$ and ${\cal O}(\sigm^2)$, 
respectively. Using the expression (\ref{expand-W}), 
the conditional mean $\cmg$ given by 
equation (\ref{dg-dm}) is calculated in the following way. 
For the numerator ${\cal A}$, we rewrites the expression (\ref{A}) to 
\begin{eqnarray}
&&{\cal A} 
= \frac{i}{2\pi}\left(\frac{\varg}{\varm}\right)^{1/2}
\int \int d\mu d\nu ~\frac{\partial}{\partial \nu} \delta_D(\nu) 
\cdot e^{-i\mu x+{\cal W(\mu,\nu)}}.
\label{cal A}
\end{eqnarray}
Integrating by part and substituting the expansion (\ref{expand-W}) 
into equation (\ref{cal A}), 
we get
\begin{eqnarray}
  {\cal A}&=&-\frac{i}{2\pi} \left(\frac{\varg}{\varm}\right)^{1/2}
  \int d\mu~ e^{-i\mu x+{\cal W}(\mu,0)}
  \left.\frac{\partial}{\partial\nu}{\cal W}(\mu,\nu)\right|_{\nu=0},
\nonumber 
\\
&=& -\frac{i}{2\pi}\left(\frac{\varg}{\varm}\right)^{1/2}
\int d\mu~ e^{-i\mu x-\mu^2/2}
\left\{1+\frac{\lm_{30}}{6}(i\mu)^3+\left(
\frac{\lm_{40}}{24}(i\mu)^4+\frac{\lm_{30}^2}{72}(i\mu)^6\right)
+{\cal O}(\sigm^3)\right\}
\nonumber \\
&&~~~~~~~~~~~~~~~~\times\left\{
-\lm_{11}\mu+i\frac{\lm_{21}}{2}(i\mu)^2+
i\frac{\lm_{31}}{6}(i\mu)^3+{\cal O}(\sigm^3)\right\}.
\nonumber
\end{eqnarray}
Keeping the terms up to ${\cal O}(\sigm^2)$, this becomes
\begin{eqnarray}
&&{\cal A}=\frac{1}{2\pi} \left(\frac{\varg}{\varm}\right)^{1/2}
\left[\lm_{11}\left(-\frac{d}{dx}\right)+
\left\{\frac{\lm_{21}}{2}\left(-\frac{d}{dx}\right)^2+
\frac{\lm_{11}\lm_{30}}{6}\left(-\frac{d}{dx}\right)^4\right\}+
\left\{\frac{\lm_{31}}{6}\left(-\frac{d}{dx}\right)^3
\right.\right.
\nonumber \\
&&+\left.\left.
\left(\frac{\lm_{11}\lm_{40}}{24}+\frac{\lm_{30}\lm_{21}}{12}\right)
\left(-\frac{d}{dx}\right)^5+\frac{\lm_{11}\lm_{30}^2}{72}
\left(-\frac{d}{dx}\right)^7\right\}+{\cal O}(\sigm^3)\right]
\int d\mu e^{-i\mu x-\mu^2/2}, 
\nonumber
\end{eqnarray}
where we replaced $(i\mu)^n$ with $\left(-\frac{d}{dx}\right)^n$. 
Performing the integration of $\mu$ and  
the differentiation with respect to $x$, we obtain the expression 
(\ref{reduced-A}). 

The calculation of the denominator in equation (\ref{dg-dm}) 
is similar to that of the numerator. Substitution of 
the expansion (\ref{expand-W}) into equation (\ref{PDF-dm}) yields 
\begin{eqnarray}
&&  {\cal P}(\dm)=\frac{1}{2\pi\sqrt{\varm}}\int d\mu~
e^{-i\mu x-\mu^2/2}\left[1+\frac{\lm_{30}}{6}(i\mu)^3+
\left\{\frac{\lm_{40}}{24}(i\mu)^4+\frac{\lm_{30}^2}{72}(i\mu)^6
\right\}+{\cal O}(\sigm^3)\right].
\nonumber
\end{eqnarray}
Repeating the same manipulation as in the above, we have 
\begin{eqnarray}
&&  {\cal P}(\dm)=\frac{1}{2\pi\sqrt{\varm}}
\left[1+\frac{\lm_{30}}{6}\left(-\frac{d}{dx}\right)^3+
\left\{\frac{\lm_{40}}{24}\left(-\frac{d}{dx}\right)^4+
\frac{\lm_{30}^2}{72}\left(-\frac{d}{dx}\right)^6
\right\}+{\cal O}(\sigm^3)\right]e^{-x^2/2}.
\nonumber
\end{eqnarray}
Using the definition of the Hermite polynomials (\ref{Hermite}), 
we obtain the expression (\ref{dm-PDF}).
%
%
\section*{Appendix B: Second order perturbation and joint moments}
%
%
%
%
When we investigate the time evolution 
of the conditional mean $\cmg$ and $\la\dm\ra|_{\dg}$ in the 
tree-level analysis, 
in addition to the biasing parameter $b$ and the cross correlation $r$,  
we need the joint moments $S_{ij}$ with $ i+j=3$: 
\begin{eqnarray}
&&  S_{30}=3\frac{\la[\dm^{(1)}]^2\dm^{(2)}\ra}{\la[\dm^{(1)}]^2\ra^2}, ~~~~
  S_{21}=\frac{\la[\dm^{(1)}]^2\dg^{(2)}\ra+
    2\la\dm^{(1)}\dm^{(2)}\dg^{(1)}\ra}{b(t)~\la[\dm^{(1)}]^2\ra^2}
\nonumber \\
&&  S_{12}=\frac{\la\dm^{(2)}[\dg^{(1)}]^2\ra+
    2\la\dm^{(1)}\dg^{(1)}\dg^{(2)}\ra}{b^2(t)~\la[\dm^{(1)}]^2\ra^2}, ~~~~
  S_{03}=3\frac{\la[\dg^{(1)}]^2\dg^{(2)}\ra}{b^3(t)~\la[\dm^{(1)}]^2\ra^2},
\label{tree-joint-moment}
\end{eqnarray}
%
%
If we evaluate these quantities for the smoothed 
density fields, the integration including the window function $W_R(x)$ 
must be computed. Although the integration in real space is difficult for 
the higher order moments, it is known that the calculation is tractable 
in the Fourier space because of the useful formulae (\cite{B94}). 
Using these formulae, we shall write down the solutions of second order 
perturbation in the Fourier space. The variables $\dm, \dg$ and $\theta$ 
are expanded as 
\begin{eqnarray}
&&  \delta_{m,g}(\x,t)=\int\frac{d^3\k}{(2\pi)^3}\hat{\delta}_{m,g}
  (\k,t)e^{-i \k\x},~~~
  \theta(\x,t)=\int\frac{d^3\k}{(2\pi)^3}\hat{\theta}(\k,t)e^{-i\k\x}.
\nonumber
\end{eqnarray}
Then the second order solutions satisfying the initial conditions are 
obtained (\cite{F84}, \cite{taruya99}):
\begin{eqnarray}
  \hat{\delta}_m^{(2)}(\k,t)&=&\int\frac{d^3\k'}{(2\pi)^3}
  \left[D^2(t)\left(\frac{6}{7}{\cal R}(\k',\k-\k')
      +\frac{1}{7}{\cal R}(\k-\k',\k')-\frac{3}{2}{\cal L}(\k',\k-\k')\right)
\right.
\nonumber\\
&&~~~~~~~~~~~~~~~~~~~~~~~~~~~~~~
\left.+\frac{3}{4}E(t){\cal L}(\k',\k-\k')\right]\hat{\Delta}_m(\k-\k')
\hat{\Delta}_m(\k'),
\label{2nd-sol-m}
\end{eqnarray}
\begin{eqnarray}
\hat{\delta}^{(2)}_g(\k,t)&=&\hat{\delta}_m^{(2)}(\k,t)-
\hat{\delta}_m^{(2)}(\k,t_i)
\nonumber\\
&+&(D(t)-1)\int\frac{d^3\k'}{(2\pi)^3}{\cal R}(\k',\k-\k')
(\hat{\Delta}_m(\k')\hat{\Delta}_g(\k-\k')
-\hat{\Delta}_m(\k')\hat{\Delta}_m(\k-\k')) 
\nonumber\\
&+&\frac{h}{6}~\int\frac{d^3\k'}{(2\pi)^3}
\left(\hat{\Delta}_g(\k')\hat{\Delta}_g(\k-\k')
  -\langle\hat{\Delta}_g^2\rangle\delta_D(\k')\right), 
\label{2nd-sol-g}
\end{eqnarray}
where
\begin{eqnarray}
&&{\cal R}(\k_1,\k_2)=1+\frac{(\k_1\cdot \k_2)}{|\k_1|^2},~~~~~
  {\cal L}(\k_1,\k_2)=1-\frac{(\k_1\cdot \k_2)^2}{|\k_1|^2|\k_2|^2}. 
\nonumber
\end{eqnarray}
The solutions (\ref{2nd-sol-m}) and (\ref{2nd-sol-g}) contain the function 
$E(t)$ which satisfies $E(t_i)=1$. This is the inhomogeneous 
solution of the following equation:
\begin{eqnarray}
&&  \ddot{E}+2H\dot{E}-\frac{3}{2}H^2\Omega E=
3H^2\Omega D^2+\frac{8}{3}\dot{D}^2.    
\nonumber
\end{eqnarray}
In Einstein-de Sitter universe, we have  
\begin{eqnarray}
&& E(t)=\frac{34}{21}D^2(t). 
\nonumber
\end{eqnarray}
It is known that the $\Omega$ and $\Lambda$ dependences of $E/D^2$ 
is extremely weak (\cite{B94}). 
Therefore, we proceed to evaluate $S_{ij}$ by replacing $E(t)$ with 
$(34/21)D^2(t)$.
Substituting the solutions (\ref{2nd-sol-m}) and (\ref{2nd-sol-g}) into 
the expression (\ref{tree-joint-moment}) and using the 
formulae below, we get the results (\ref{s30})-(\ref{s03}): 
\begin{eqnarray}
  \label{formula1}
&& \int\int\frac{d^3\k_1}{(2\pi)^3}\frac{d^3\k_2}{(2\pi)^3}
  {\cal R}(\k_1,\k_2)P(k_1)P(k_2)\tilde{W}_R(|\k_1+\k_2|R)
  \tilde{W}(k_1R)\tilde{W}(k_2R)=\sigma_0^4(R)
  \left(1-\frac{\gamma}{6}\right),
\nonumber\\
&& \int\int\frac{d^3\k_1}{(2\pi)^3}\frac{d^3\k_2}{(2\pi)^3}
  {\cal L}(\k_1,\k_2)P(k_1)P(k_2)\tilde{W}_R(|\k_1+\k_2|R)
  \tilde{W}(k_1R)\tilde{W}(k_2R)=\frac{2}{3}\sigma_0^4(R), 
\nonumber
\end{eqnarray}
where 
\begin{eqnarray}
&&\gamma=-\frac{d}{d(\log{R})}[\log{\sigma_0^2(R)}].
\nonumber
\end{eqnarray}
$\tilde{W}_R(kR)$ and $\sigma_0$ are the window function 
in the Fourier space and the initial variance of the total mass 
defined by equation (\ref{init-var}), respectively.  
\clearpage
%
%
%

%
%
%
%
%
%
%
%
%
%
\newpage
%
%
%
%
\section*{Figure Caption}
\begin{description}
%
%
%
\item[Fig.1] 
The time evolution of the coefficients $c_1$ and $c_2$ is evaluated 
as a function of the expansion factor $a(t)$ in the case of the Einstein-de Sitter 
universe.  For each figure, 
the solid line shows the coefficient $c_1$. The long-dashed, the short-dashed 
and the dotted lines in Fig.1a and 1b denote the 
time evolution of $c_2$ with the spectral index $n=-2$, $-1.5$ and 
$-1$, respectively. The dashed line in Fig.1c also shows the coefficient 
$c_2$, but we set the spectral index $n=-1.41$. The initial parameters for 
each figure are as follows: (a) $b_0=2.0$, $r_0=0.8$ and $h=3.0$; 
(b) $b_0=2.0$, $r_0=0.1$ and $h=3.0$; (c) $b_0=4.63$, $r_0=0.2$ and $h=6.96$. 
Note that the parameters in Fig.1c gives the same skewness and the bi-spectrum 
as obtained from the Lick catalog data at $a=4$ if the universe is 
assumed to be the Einstein-de Sitter universe (see section 
\ref{subsec: evolution-cmg}). 
\item[Fig.2] 
The time evolution of the galaxy-mass density relation 
  $\cmg$ is plotted in $(\dg,\dm)$-space. The initial parameters are 
  the same as Fig.1c. Each line represents the snapshot evaluated at 
  $a=1$ ({\it solid line}), $a=2$ ({\it long-dashed line}), 
  $a=4$ ({\it short-dashed line}) and $a=8$ ({\it dotted line}). 
%
%
%
\item[Fig.3] 
The time evolution of the skewness for the galaxy distribution is shown 
as a function of the expansion factor $a$ in the Einstein-de Sitter universe. 
The solid line is the joint moment $S_{03}$ 
multiplied by the factor $1/b(t)$, which is identical to the correct 
value of the skewness $S_{3,g}$. The dashed line denotes the skewness 
$\tilde{S}_{3,g}$ deduced from the galaxy-mass density relation 
(\ref{dg-dm3}), which is evaluated from equation (\ref{tilde-skewness}). 
Fig.3a and 3b respectively have the same initial parameters 
as Fig.1a and Fig.1b, except for the spectral index specified as $n=-2$. 
As for Fig.3c, the initial parameters are the same parameters 
as Fig.1c, which gives the same results as obtained from the Lick catalog data 
if we identify the initial time $a(t)=1$ with the redshift parameter $z=3$ 
in the Einstein-de Sitter universe. 
\end{description}
%
%
%
%
%
%
%
%
%
\newpage

\epsfxsize=\hsize
\epsfbox{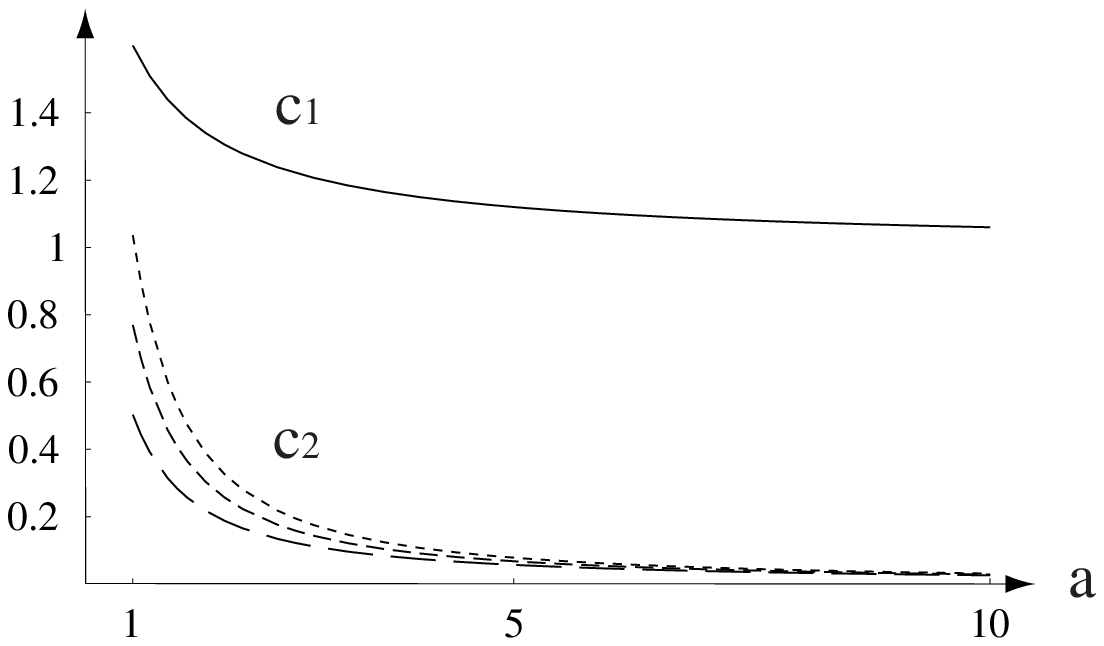}     
\vspace{2.5cm}
\begin{center}
{\large Fig.1a}
\end{center}

\epsfxsize=\hsize
\epsfbox{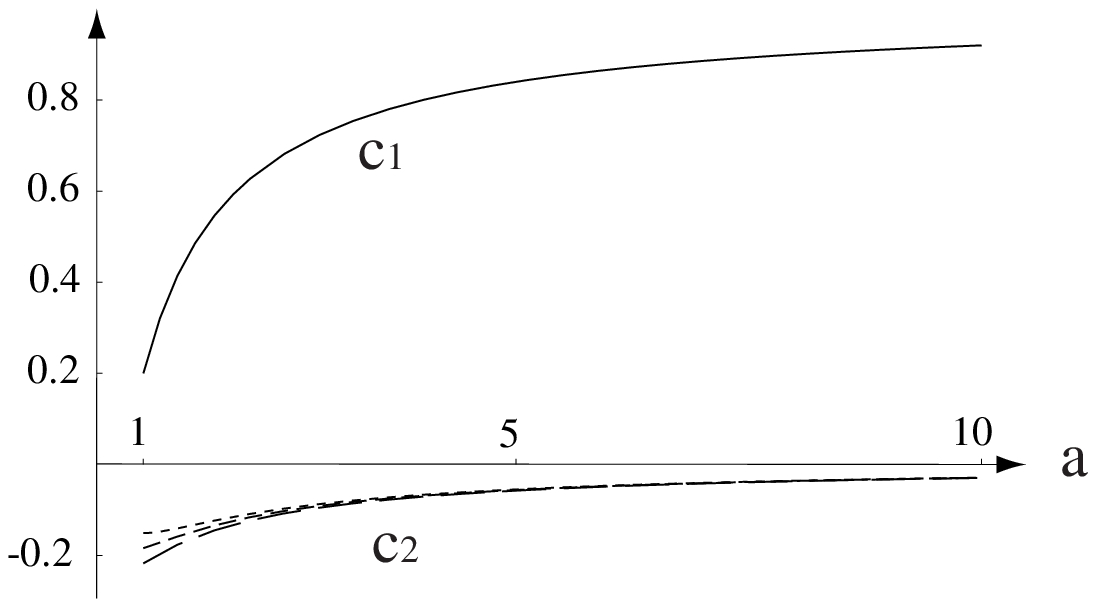}     
\vspace{2.5cm}
\begin{center}
{\large Fig.1b}
\end{center}

\epsfxsize=\hsize
\epsfbox{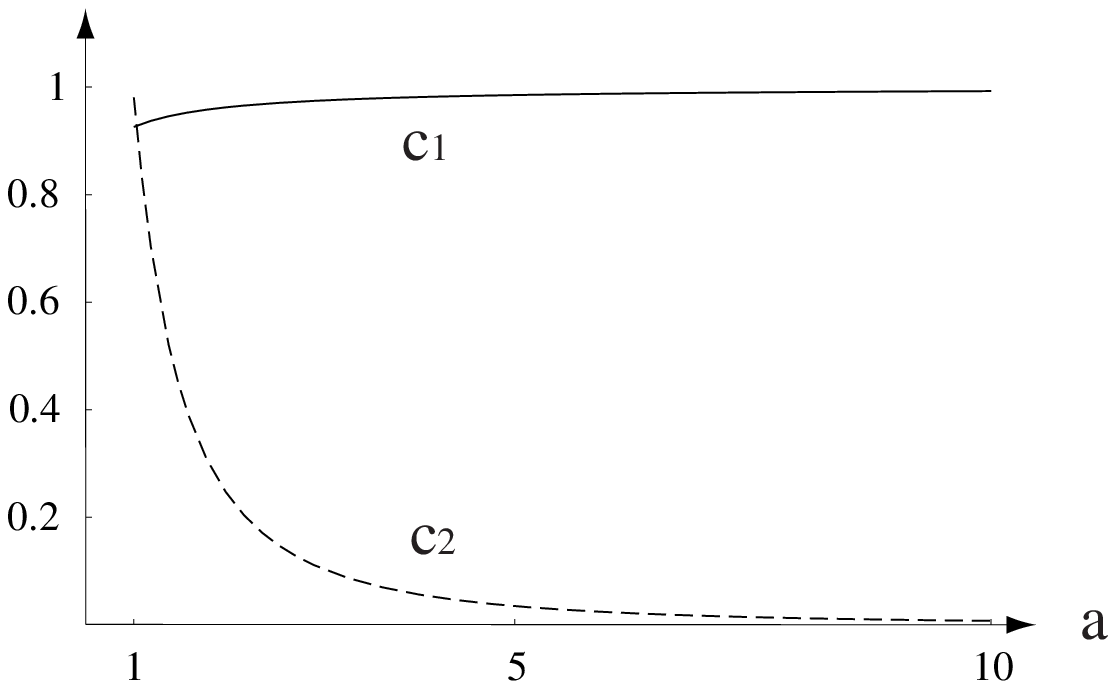}     
\vspace{2.5cm}
\begin{center}
{\large Fig.1c}
\end{center}

\epsfxsize=0.9\hsize
\epsfbox{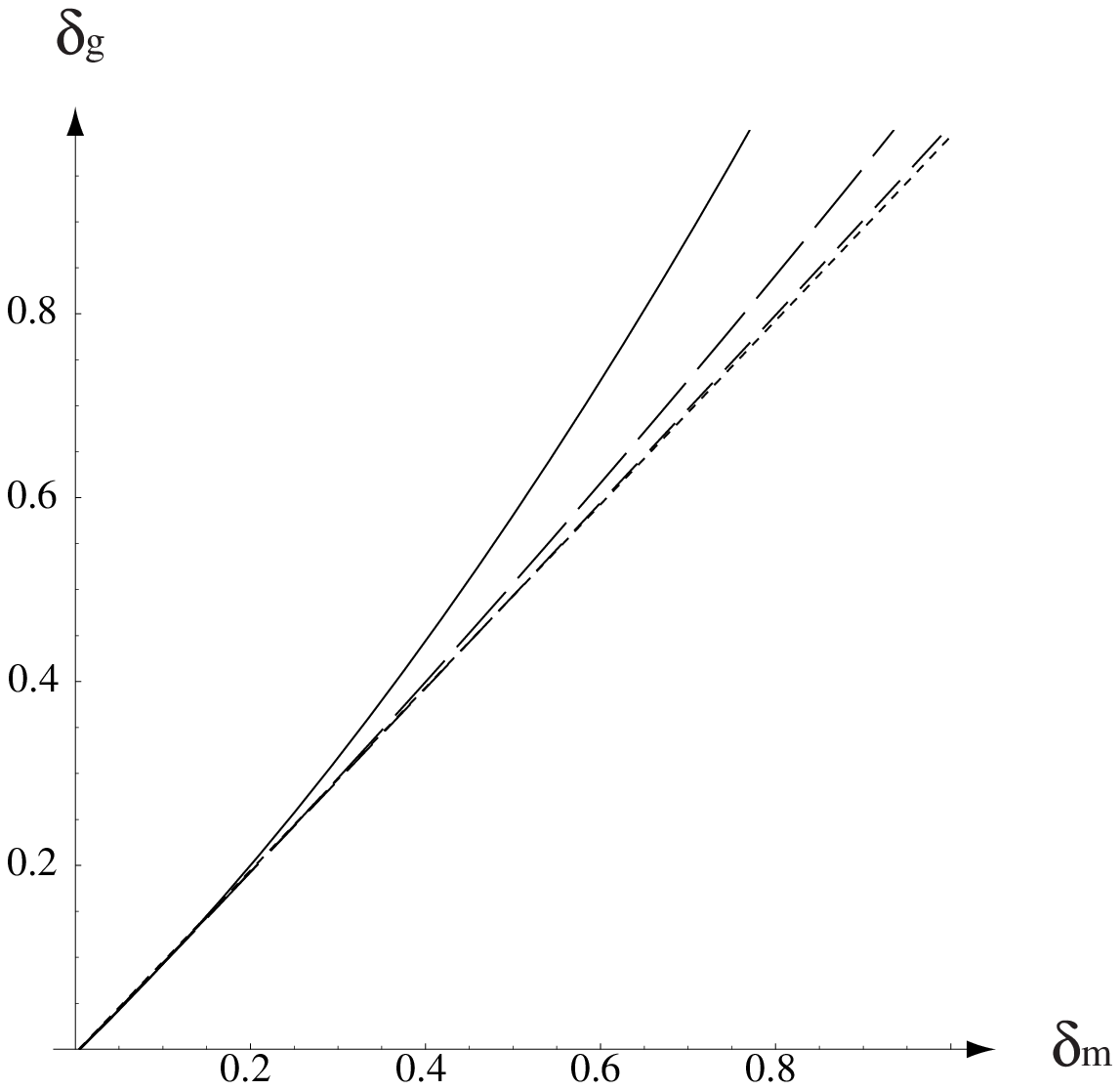}     
\vspace{2.5cm}
\begin{center}
{\large Fig.2}
\end{center}

\epsfxsize=\hsize
\epsfbox{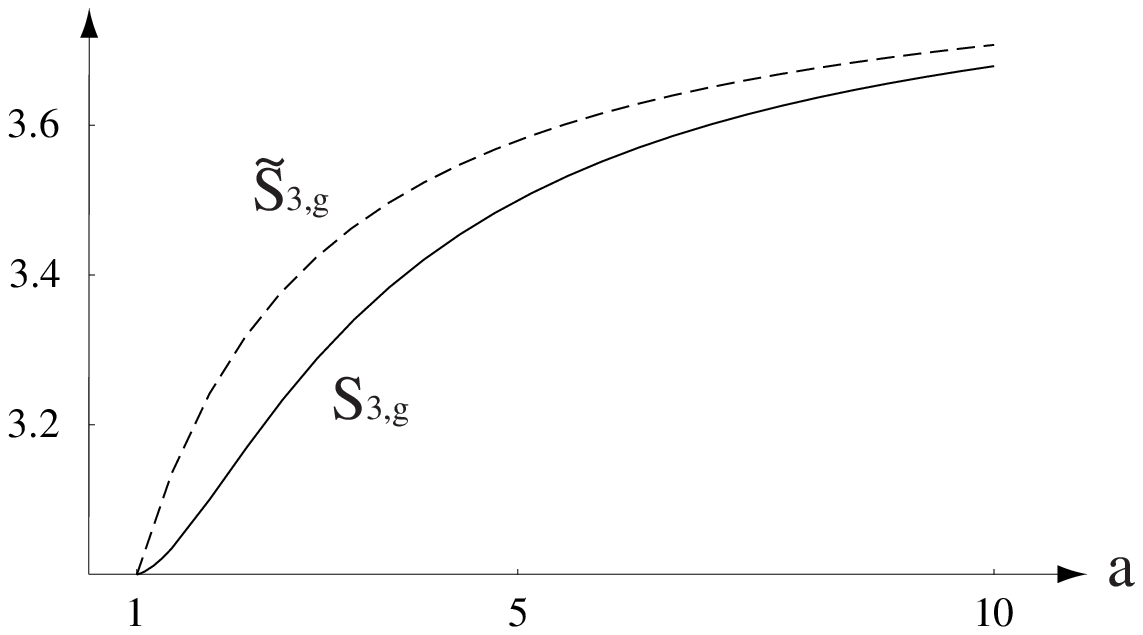}     
\vspace{2.5cm}
\begin{center}
{\large Fig.3a}
\end{center}

\epsfxsize=\hsize
\epsfbox{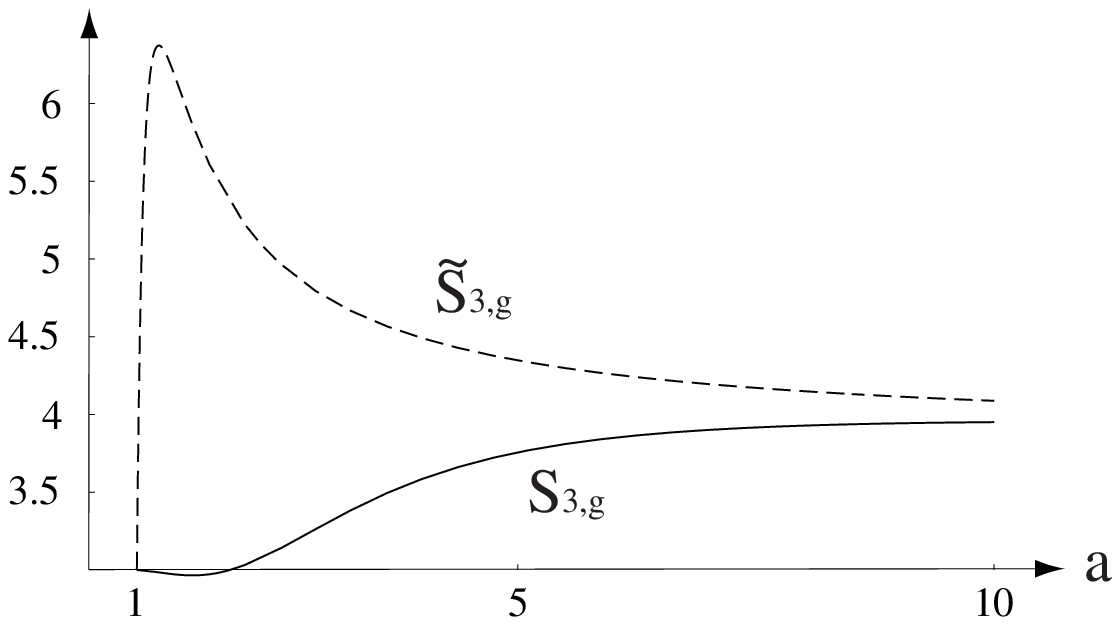}     
\vspace{2.5cm}
\begin{center}
{\large Fig.3b}
\end{center}

\epsfxsize=\hsize
\epsfbox{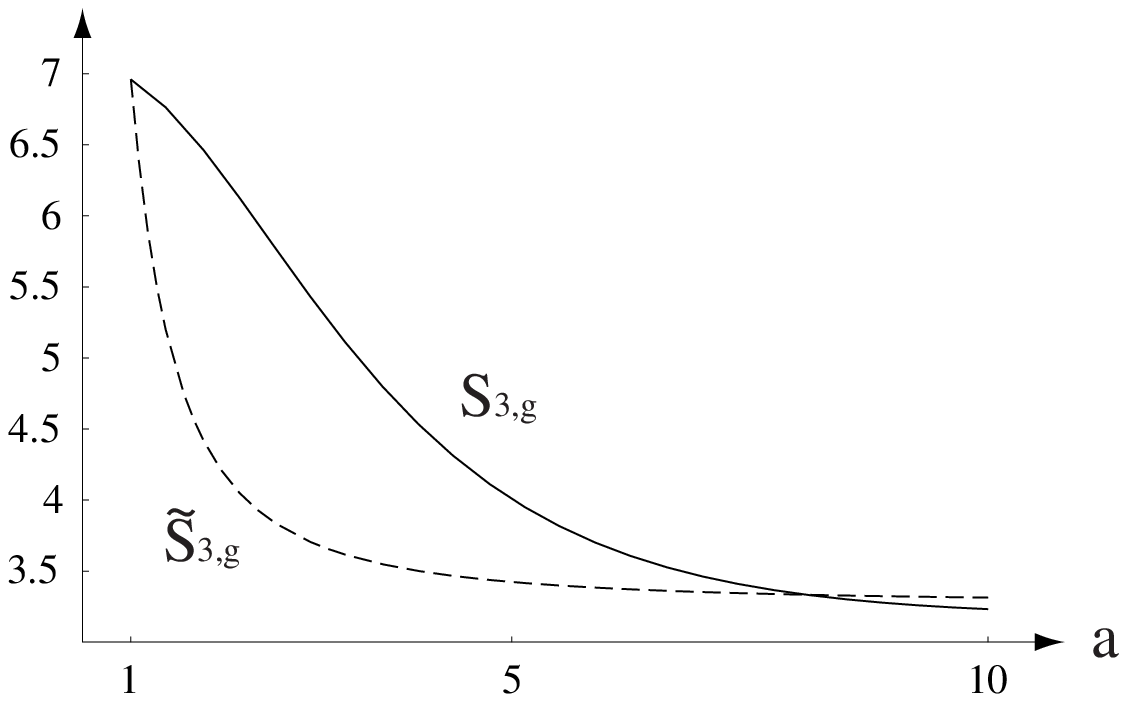}     
\vspace{2.5cm}
\begin{center}
{\large Fig.3c}
\end{center}
\end{document}